\documentclass{9-SNMAM}

\usepackage{amssymb}
\usepackage{graphicx}
\usepackage{multirow}
\usepackage[table,xcdraw]{xcolor}
\usepackage{url}

\newcommand{\keywords}[1]{\par\addvspace\baselineskip
\noindent\keywordname\enspace\ignorespaces#1}
\begin{document}
\mainmatter 
\title{Maximising Influence Spread in Complex Networks by Utilising Community-based Driver Nodes as Seeds}
\titlerunning{Maximising Influence} 
\author{Abida Sadaf\inst{1} \and Luke Mathieson\inst{1}
Piotr Br\'{o}dka\inst{2} \and Katarzyna Musial\inst{1}}

\authorrunning{Abida Sadaf et al.} 
\institute{Complex Adaptive Systems Lab, School of Computer Science,  University of Technology Sydney, Sydney, Australia\\
\url{http://www.uts.edu.au}
Department of Artificial Intelligence, Wroclaw University of Science and Technology, Wroclaw, Poland\\
\url{https://pwr.edu.pl/}}

\setlength{\textfloatsep}{10pt plus 1.0pt minus 2.0pt}
\setlength{\floatsep}{6pt plus 1.0pt minus 1.0pt}
\setlength{\intextsep}{6pt plus 1.0pt minus 1.0pt}
\setlength{\parskip}{0pt}

\maketitle
\begin{abstract}
Finding a small subset of influential nodes to maximise influence spread in a complex network is an active area of research. Different methods have been proposed in the past to identify a set of seed nodes that can help achieve a faster spread of influence in the network. 
This paper combines driver node selection methods from the field of network control, with the divide-and-conquer approach of using community structure to guide the selection of candidate seed nodes from the driver nodes of the communities.

The use of driver nodes in communities as seed nodes is a comparatively new idea. We identify communities of synthetic (i.e., Random, Small-World and Scale-Free) networks as well as twenty-two real-world social networks. Driver nodes from those communities are then ranked according to a range of common centrality measures. We compare the influence spreading power of these seed sets to the results of selecting driver nodes at a global level. We show that in both synthetic and real networks, exploiting community structure enhances the power of the resulting seed sets. 

\keywords{Influence, Complex Network, Social Networks, Seed Selection Methods, Driver Nodes, Communities}
\end{abstract}
\section{Introduction}
Due to the prevailing use of online social networking sites, social networks are very much a hot topic in network science. Nowadays, we have a good understanding of network structures and attention has shifted more towards their prediction, influence, and control. Full control of social networks is very hard to achieve due to their varying structures, dynamics, and the complexities of human behaviour. This study looks into how driver nodes, which enable complex network control, can be used in the context of influence spread in the social network space. We use driver nodes at both the global and community level to `divide and conquer' the time-consuming problem of driver node identification.
Until recently, we did not know if and how the structure of social networks correlated with the number of driver nodes required to control the network~\cite{sadaf2021insight}. As driver nodes play a key role in achieving control of a complex network, identifying them and studying their correlation with network structure measures can bring valuable insights, such as what network structures are easier to control, and how we can alter the structure in our favour to achieve the maximum control over the network. Our previous work~\cite{sadaf2021insight} determines the relationship between some global network structure measures and the number of driver nodes. This study builds an understanding of how global network profiles of synthetic (random, small-world, scale-free) and real social networks influence the number of driver nodes needed for control. It focuses on global structural measures such as network density and how it can play an important role in determining the size of a suitable set of driver nodes. Our results show that as density increases in networks with structures exhibited by random, small world and scale free networks, the number of driver nodes tends to decrease.  
In this work we explore the potential that exploiting local structures (in this study we focus on communities) can offer in developing control of, and influencing, the network. Finding communities in a social network is itself a difficult task due to both dynamic and combinatorial factors~\cite{sathiyakumari2016community}.

This study explores the possibility of using community structure in social networks to reduce the cost of identifying driver nodes, and whether this remains a feasible approach for network control and influence spread methods.

Our main research questions for this work are stated as follows: 
\begin{enumerate}
    \item How can we rank driver nodes within communities to identify an optimal subset of driver nodes for use as seed nodes? 
    \item How quickly does influence spread from seed nodes chosen using driver node selection methods at the community level?
    \item Does the percentage of influenced nodes increases or decreases when using driver node based seed selection methods in communities as compared to driver node based seed selection methods in the network as a whole, for both synthetic and real data?
    \item How does the network structure (of synthetic or real networks) impact the percentage of nodes influenced with each method?
\end{enumerate}
This paper contains the following sections: Section~\ref{background} describes related work and the main research challenge that is the focus of this study. 
Sections~\ref{methodology} and~\ref{resultsandanalysis} describe (i) the research methodology in detail and (ii) include results and analysis of the experiments performed respectively. Finally, the conclusions drawn from the experiments and future work are discussed in Section~\ref{conclusion&futurework}.

\section{Related Work}\label{background}
The Influence Maximisation problem aims at discovering an influential set of nodes that can influence the highest number of nodes in social networks in the shortest possible time. A set of these nodes can be used to propagate influence in terms of social media news, advertising, etc. Several algorithms have been proposed to solve the influence maximisation problem that identify a set of nodes that is highly influential as compared to other nodes. For example Basic Greedy~\cite{kempe2003maximizing}, CELF~\cite{leskovec2005graphs}, CELF++~\cite{goyal2011celf++}, Static Greedy~\cite{cheng2013staticgreedy}, Nguyen’s Method~\cite{nguyen2013budgeted}, Brog et al.'s Method~\cite{borgs2014maximizing},SKIM~\cite{cohen2014sketch}, TIM+~\cite{tang2014influence}, IMM~\cite{tang2015influence}, Stop and Stare~\cite{nguyen2016stop}, Zohu et al.’s Method~\cite{zhu2017emotional} and BCT~\cite{nguyen2017billion} are some of those algorithms.  Many algorithms have high run times when identifying a set of nodes to diffuse the influence through a social network, therefore there is a need to work on exploring different types of nodes if those can work towards achieving the high influence~\cite{kazemzadeh2022influence}. 
The problem of influence maximisation has high relevancy to the spreading of information on networks. The two most common network-based models are Independent Cascade model~\cite{kempe2003maximizing} and Threshold models~\cite{granovetter1978threshold}.
In one of the previously proposed framework, the possible seed set has been identified by analysing the properties of the community structures in the networks. The CIM algorithm (i.e. Community-Based Influence Maximisation), utilises hierarchical clustering to detect communities from the networks and then uses the information of community structures to identify the possible seed nodes candidates, and at the end the final seed set is selected from the candidate seed nodes~\cite{chen2014cim}. 
From the previous work such as~\cite{chen2014cim} and~\cite{kazemzadeh2022influence}, we can see, that by detecting communities and then selecting seed nodes from those communities can be an effective strategy to maximise influence. 
 
From previous study~\cite{sadaf2021insight}, following main results were achieved, which are the basis for further new experiments in this current research work.
    \begin{itemize}
        \item Correlation between network density and number of driver nodes: For this purpose, network densities and number of driver nodes in those networks are plotted against each other to see the increase/decrease in number of driver nodes with the increase/decrease in the densities of the networks.
        \item Structural measures and density of driver nodes: In this step a comparison of structural measures like (Betweenness Centrality, Closeness Centrality, Nodes, Edges, Eigenvector Centrality and Clustering Coefficient) is presented with the density of number of driver nodes. Density of number of driver nodes is defined as total number of driver nodes divided by total number of nodes in the network.
    \end{itemize}
In our proposed methods, we utilise driver nodes within the communities of networks for the influence spread using Linear Threshold Model. To make the driver nodes more influential, we propose different ranking mechanisms to see the number of nodes influenced after a certain time with a certain percentage of seed nodes in synthetic as well as real networks. The detail of network datasets has been presented in the later sections.
We explain our method to select seed nodes from the communities in the next section.

\section{Methodology}\label{methodology}

This work springs from the question, whether network control methods, in particular driver node selection, can be used to improve seed selection in influence models.

This prompts two possible approaches: (i) using driver nodes selected from the network as a whole, and (ii) using driver nodes selected at the community level as seeds.
For all experiments, we used the Linear Threshold Model to model influence propagation. We used a set threshold of 0.5 for the network diffusion model. We have previously observed that a threshold value of at least 0.4 accelerates influence propagation~\cite{chen2014cim}.
\subsection{Datasets Description}
To enable comprehensive and robust testing of the proposed approaches, both generated and real-world social networks have been used. Following is a brief description of networks used in the experiments.
    \begin{enumerate}
        \item Generated Networks: we generated random, small-world and scale free networks from network size of (100, 200, 300, 400, 500) nodes. For each network size (from 100 to 500), we generated networks with increasing density, to the maximum density of 1. A total of 720 networks were generated~\cite{sadaf2021insight}. 
        \item Social Networks: we use 22 real-world social networks of varying size, the number of nodes and number of edges are presented in Table~\ref{table:Gainoverothermethodssocialnetworks}. The networks are available for download at SNAP\footnote{http://snap.stanford.edu/}. 
    \end{enumerate}
\subsection{Influence spread using global driver nodes as seeds}
The first experiment focuses on the seed selection process from the global perspective. Driver nodes are selected from the network as a whole, ranked, and finally used as seeds in the influence process. The below described approach has been proposed in~\cite{sadaf2022bridge}. As it outperforms other state-of-the art ranking methods, it serves in this study as a benchmark to show a difference between global- and local-level seed selection methods. The steps are as follows:
\begin{enumerate}
    \item Minimum Dominating Set method~\cite{nacher2012dominating} has been used to identify the number of driver nodes from the networks. More detail of this process can be found in~\cite{sadaf2021insight}. DMS has been found by using greedy algorithm. At start, the dominating set is empty. Then in each iteration of the algorithm, a vertex is added to the set such that it covers the maximum number of previously uncovered vertices. Then, if more than one vertex fulfils this criteria, the vertex is added randomly among the set of nominated vertices~\cite{sanchis2002experimental}. 
    
    \item We ranked the nodes using different ranking mechanisms. The goal was to achieve an efficient set of nodes as seeds that can achieve maximum or full influence more quickly.
    The ranking mechanisms used are: Random, Degree Centrality, Closeness Centrality, Betweenness Centrality, Kempe Ranking, Degree-Closeness-Betweenness. We tested various seed set sizes: 1\%, 10\%, 20\%, 30\%, 40\% and 50\% of all detected driver nodes ranked these methods. In each of the methods, the driver nodes are ranked based on the following measures:
    \begin{itemize}
        \item In Random (Driver Random -- DR) we ranked the driver nodes randomly.
        \item In Degree seed selection (DD) we ranked the driver nodes based on their degree in descending order.
        \item For Closeness Centrality based seed selection method (Driver Closeness -- DC), we ranked the nodes on the basis of their closeness centrality in descending order.
        \item For Betweenness Centrality based seed selection method (Driver Betweenness -- DB), we ranked the nodes on the basis of their betweenness centrality in descending order.
        \item For Degree-Closeness-Betweenness method (Driver Degree Closeness Betweenness -- DDCB), we ranked (in descending order) the driver nodes on the basis of the average of degree, closeness and betweenness centralities of each driver nodes.
        \item For Kempe ranking (Driver Kempe -- DK),  we start by spreading influence through all the driver nodes as seed nodes. So we calculate the total number of nodes influenced by each driver node already in the seed set, and then rank them in descending order. After ranking, we select a percentage of nodes that are required for a seed set. 
        \item Linear Threshold Model (LTM) has been implemented for influence spread process. In LTM the idea is that a node becomes active if a sufficient part of its neighbourhood is active. Each node $u$ has a threshold $t \in[0, 1]$. The threshold represents the fraction of neighbours of $u$ that must be active in order for $u$ to become active. At the beginning of the process, a small percentage of nodes (seeds) is set as active in order to start the process. In the next steps a node becomes active if the fraction of its active neighbours is greater than its threshold, and the whole process stops when no node is activated in the current step~\cite{d2016influence}.
    \end{itemize} 
\end{enumerate}
\subsection{Influence spread using local driver nodes as seeds}
The second experiment employs a new strategy: first identify communities in the network, and then identify driver nodes on a per-community basis. 

\begin{figure}
\centering
\includegraphics[height=7cm]{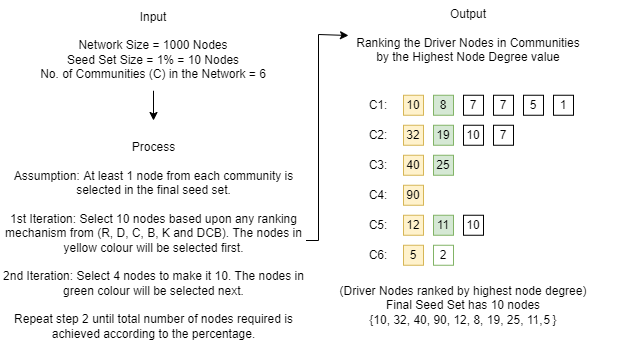}
\caption{An example showing the process for selecting seed nodes set from the driver nodes identified in network communities}
\label{fig:ExampleSelectionofSeedSetfromCommunities}
\end{figure}

Once driver nodes for each community are identified, they are then ranked using the same ranking mechanisms as in the first experiment, with seed sets chosen to cover all communities (detailed below).
In detail, the approach is as follows:
\begin{enumerate}
\item Firstly, communities are identified in the network. This was done using Girvan-Newman algorithm~\cite{girvan2002community}. The Girvan–Newman algorithm detects communities by progressively removing edges from the original graph in order of the highest betweenness centrality.
\item  Within each community, candidate driver nodes were identified using the Minimum Dominating Set~\cite{nacher2012dominating} approach as used with the whole network.
Correlation between community densities and number of driver nodes is found by obtaining densities of the communities and identifying number of driver nodes in those communities by MDS method. Difference (Diff.) between total number of driver nodes identified in overall networks (NDN) as compared to the number of driver nodes found in communities of those networks (NDNC) is also obtained. The Diff. tells us, the significance of identifying driver nodes within communities, like following a divide and conquer approach.
\item To rank the nodes, we introduce a multi-round selection process. This process effectively ranks driver nodes within each community according to the ranking criterion, then selects one node per community per round, in the order given by the ranking, until the total percentage to be chosen is reached. This is perhaps better explained by the following example, illustrated in Figure~\ref{fig:ExampleSelectionofSeedSetfromCommunities}. Consider a network with 1,000 nodes and 6 communities. Select a ranking method, in this case the node degree. Choose a target percentage of nodes to use as seed nodes, 1\% in the example. Now, in order to choose 10 nodes from the driver nodes detected in the communities, we select 6 nodes at first -- the highest degree node from each community, marked in yellow in the figure. In the second round, we can select at most 4 nodes to reach the target of 10 -- from each community, we take the node with the second-highest node degree and rank these nodes according to their degrees and take the 4 nodes with the highest degree.  We choose the same ranking mechanism for all the community based driver nodes seed selection methods i.e., the highest node degree, apart from the original ranking that is different in each technique as explained previously.
\item Influence spread in the overall network using Driver Based Seed Selection Methods is done by following a series of steps. Starting from identification of driver nodes from the networks, ranking of driver nodes based upon Random, Node Degree, Closeness Centrality, Betweenness Centrality, Kempe Ranking, Degree-Closeness-Betweenness Centralities combined. After ranking of driver nodes, we selected our seed set on the basis of percentage of nodes from that set. We run our LTM for different seed sets, namely for example 1\%, 10\%, 20\%, 30\%, 40\% and 50\%.
\item Influence spread through Driver Nodes in communities of Networks is done by identifying driver nodes in communities. However, there was a challenge of getting the ultimate seed set that has representation from all the communities of the network. For this purpose, we devised our ranking approach that makes sure that at least one driver node is selected from each community of the network to make sure that the nodes in those communities can also be part of the influence process. For each of the driver based seed selection methods, we used one unified approach to further rank the nodes so that we are able to select at least one node from each of the communities.
\end{enumerate}

\section{Results and Analysis}\label{resultsandanalysis}

\begin{figure}[htb]
    \centering
    \includegraphics[height=8cm]{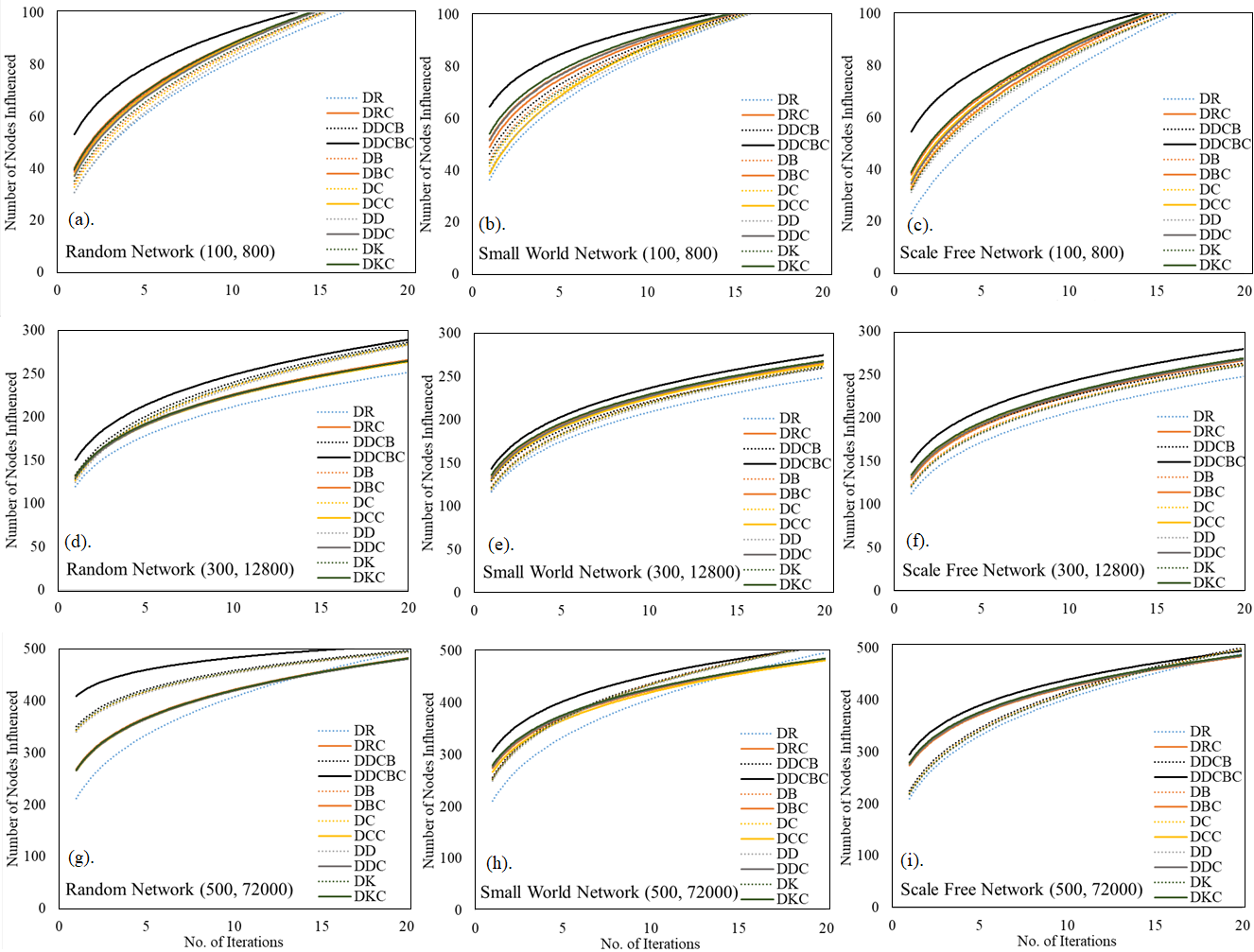}
    \caption{Number of Nodes Influenced in Random, Small-World and Scale-Free Networks: when the number of nodes (N) is 100 and the number of edges (E) is 800 (Figures a, b and c); when N is 300 and E is 12800 (Figures d, e and f); when N is 500 and E is 72000 (Figures g, h and i). A Comparison of all methods for 20 iterations when the seed size is 1\% is presented.}
    \label{fig:RSWSF-20Iterations100-300-500-Nodes}
\end{figure}
\sloppypar
Six novel network level seed selection methods (i.e. Driver-Random (DR), Driver-Degree (DD), Driver-Closeness (DC), Driver-Betweenness (DB), Driver-Kempe (DK) and Driver-Degree-Closeness-Betweenness (DDCB)) have been proposed and tested on synthetic and real world networks before in~\cite{sadaf2022bridge} and the results show that those methods outperform their non-driver based counterparts. In this study, we use those methods but instead of selecting driver nodes from the global network, we propose a local approach where driver nodes are identified within the networks' communities. We name the new methods by adding C (for community) to the previously proposed methods (i.e, DRC - Driver-Random-Community, DDC - Driver-Degree-Community, DCC - Driver-Closeness-Community, DBC - Driver-Betweenness-Community, DKC - Driver-Kempe-Community and DDCBC - Driver-Degree-Closeness-Betweenness-Community). Below, we compare community based driver seed selection methods to network based driver seed selection methods.
\subsection{Results From Generated Networks}
This section covers the results and analysis of the experiments performed on generated networks.
\subsubsection{What is the speed and reach of the influence spread?}
First, we compare the percentage of nodes influenced for global-level driver based seed selection methods and local-level (community) driver based seed selection methods. We perform the analysis iteration by iteration to see which seed selection methods enable to achieve the highest coverage the fastest.

In Figure~\ref{fig:RSWSF-20Iterations100-300-500-Nodes}, we can see trend-lines for all the seed selection methods (when seed set size is 1\% of all the driver nodes) for random, small-world and scale-free networks. DDCBC method outperforms other methods in almost all the experimented cases. We can see a `head-start' in the trend-line of DDCBC (represented in black colour) for all the networks when number of nodes in the network is 100 and number of edges is 800. This means that in only few iterations, DDCBC enables to influence more nodes than in the case of other seed selection methods.

\begin{figure}[htb]
    \centering
    \includegraphics{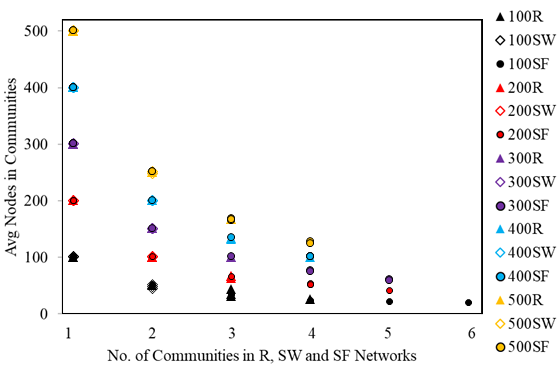}
    \caption{Average Number of Nodes in Communities of Random, Small-World and Scale-Free Networks versus number of communities in those networks. Legend shows the Number of Nodes in communities of generated networks i.e. Random (R), Small-World (SW) and Scale-Free (SF).}
    \label{fig:AvgNodesCommunities-RSWSF}
\end{figure}

Results in Figure~\ref{fig:AvgNodesCommunities-RSWSF} show that when the network is of small size, and density is approximately equal to $0.6$, the influence spreads faster when using driver-community based seed selection methods than when the global-level driver based methods are employed. 
If we look at Figure~\ref{fig:AvgNodesCommunities-RSWSF}, the network of smaller densities (i.e. $0.4$), where number of nodes is 300 and number of edges is 2,800, the difference between the global-level driver based methods and community-level driver based methods is not so big. But we do see a gap between DDCBC method and other methods. Which tells us that, so far, DDCBC ranking of driver nodes in communities is working better than when we are using driver nodes of communities as seed nodes.

Although the comparison is done on a very small size of seed set (1\% of all driver nodes), in DDCBC, we still achieve more influence earlier in the spreading process when using community-level driver based methods. It also gives us another insight regarding larger networks, their structures and densities, and how those are connected to spreading influence. We see that the spread is faster when density is higher than $0.5$ as in the case of networks presented in the Figure~\ref{fig:RSWSF-20Iterations100-300-500-Nodes} (network with 500 nodes and 72,000 edges). We can see that in those cases, the driver-community based method DRC, DDC, DBC, DKC and DDCBC outperforms their counterpart methods DR, DD, DB, DK and DDCB. 

Based upon these observations, we conclude it does not matter which type of network it is, as long as its density is higher than $0.5$ it will respond to the community-based seed selection methods better and the spread will be faster. Also, regardless of the network density, community-based method -- DDCBC -- outperforms all other methods Figure~\ref{fig:RSWSF-20Iterations100-300-500-Nodes}(a-f). This holds true for all the other settings as well. As when we have different edges for 100, 200, 300, 400 and 500 nodes networks.

\subsubsection{How much advantage do community—level driver based seed selection methods give?}
Given a number of iterations $n$ and a method $X$, let $N^{infl}_{n}(X)$ denote the number of nodes influenced using the method $X$ after $n$ iterations. The Percentage Gain of method $A$ over method $B$ after $n$ iterations is then given by:
\begin{equation}\label{eqn:percent_gain}
\frac{N^{infl}_{n}(A) - N^{infl}_{n}(B)}{N} \times 100
\end{equation}
where $N$ is the number of nodes in the network.

Table~\ref{table:Gainoverothermethodsgeneratednetworks} shows the percentage gain of the DDCBC method over the global-level driver based methods. We represent only driver based methods (i.e. DR, DB, DC, DD, DK and DDCB), as the gain is higher over these methods as compared to other driver-community based methods (i.e. DRC, DBC, DCC, DDC and DKC) as well as they are our baseline for this study. Percentage gain is calculated by knowing the maximum number of nodes influenced after $20$ iterations when seed size is 1\%.

From Table~\ref{table:Gainoverothermethodsgeneratednetworks} we can see the maximum gain in when the average density of the communities of the network is greater than 0.5. When the density reaches 1 all the methods perform very similar as spread in fully connected network behaves in a very similar way regardless of applied seed selection method. This highlights our previous point that density of network plays an important part in how effective a network is going to respond to the influence spread process. We can see the highest gain for DDCBC method in random networks, but DDCBC outperforms all global-level driver based methods in all the networks, except for the networks with densities equal or very close to 1.

From Figure~\ref{fig:AvgNodesCommunities-RSWSF}, we can see the number of average nodes in communities versus the total number of communities in Random, Small-World and Scale-Free networks.  The denser the network, the fewer communities we have, and those communities are denser than the previous ones. Hence, due to increase in community density, we see the higher percent gain in DDCBC method. The number of nodes influenced by DDCBC method increases, when there are fewer communities. Because when number of communities are less, they tend to be denser, hence the increase in number of nodes influenced. We see the difference in number of nodes influenced in DDCBC method which is bigger than compared to other methods. 

\begin{table}[ht]
\centering
\caption{\label{table:Gainoverothermethodsgeneratednetworks}A percentage gain table shows the percentage gain of DDCBC method over other seed selection methods in influencing the nodes in Random, Small-World and Scale-Free networks when the seed set size is 1\% after 20 iterations. $N$ is number of nodes, $E$ is number of edges, $C$ is number of communities and $CD$ is average community density.}
\begin{tabular}{|r|r|r|r|llllll|llllll|llllll|}
\hline
                       & \multicolumn{1}{c|}{}                    & \multicolumn{1}{c|}{}                    & \multicolumn{1}{c|}{CD} & \multicolumn{6}{c|}{Random Networks}                                                                                                                                                                                                                                                 & \multicolumn{6}{c|}{Small-World Networks}                                                                                                                                                                                                                                            & \multicolumn{6}{c|}{Scale-Free Networks}                                                                                                                                                                                                                                             \\
\hline
\end{tabular}
\end{table}
\begin{table}[ht]
\centering
\caption{\label{table:Gainoverothermethodssocialnetworks}A percentage gain table shows the percentage gain of DDCBC method over other seed selection methods in influencing the nodes of the social networks. Average Community Densities of the networks are as follows: FB (0.06±0.02), ZKC (0.32±0.4), Twitter (0.00029±0.05), Diggs (0.00008±0.007), Youtube (0.000012±0.04), Ego (0.00034±0.05), LC (0.007±0.032), LF (0.0073±0.09), PF (0.015±0.54), MFb (0.001±0.43), DHR (0.00085±0.21), DRO (0.0005±0.4), DHU (0.0004±0.63), MG (0.0011±0.03), L (0.0019±0.54), FbAR (0.0014±0.03), FbA (0.0015±0.09), FbG (0.0075±0.05), FbN (0.0013±0.003), FbP (0.0049±0.003), FbPF (0.004±0.032) and Fbt (0.0051±0.05)}
\begin{tabular}{|r|r|r|c|ccccccccccc|}
\hline  \multicolumn{1}{|c|}{} & \multicolumn{1}{c|}{}                    & \multicolumn{1}{c|}{}                    &                                   & \multicolumn{11}{c|}{Seed Selection Methods (20\% of all nodes)}                                                                                                                                                                                                                                                                                                                                                                                                                                                                                                                                  \\ \cline{5-15} 
\multicolumn{1}{|c|}{\multirow{-2}{*}{N}} & \multicolumn{1}{c|}{\multirow{-2}{*}{E}} & \multicolumn{1}{c|}{\multirow{-2}{*}{C}} & \multirow{-2}{*}{Networks} & \multicolumn{1}{c|}{DR}                            & \multicolumn{1}{c|}{DD}                            & \multicolumn{1}{c|}{DC}                            & \multicolumn{1}{c|}{DB}                            & \multicolumn{1}{c|}{DDCB}                          & \multicolumn{1}{c|}{DK}                            & \multicolumn{1}{c|}{DRC}                           & \multicolumn{1}{c|}{DDC}                           & \multicolumn{1}{c|}{DCC}                           & \multicolumn{1}{c|}{DBC}                           & DKC                                  \\ \hline
4039                                      & 88234                                    & 180                                      & FB                                & \multicolumn{1}{r|}{\cellcolor[HTML]{93BC77}28.68} & \multicolumn{1}{r|}{\cellcolor[HTML]{9FC784}25.03} & \multicolumn{1}{r|}{\cellcolor[HTML]{A0C784}24.94} & \multicolumn{1}{r|}{\cellcolor[HTML]{9CC481}25.94} & \multicolumn{1}{r|}{\cellcolor[HTML]{9FC783}25.15} & \multicolumn{1}{r|}{\cellcolor[HTML]{A1C985}24.59} & \multicolumn{1}{r|}{\cellcolor[HTML]{AAD18F}21.59} & \multicolumn{1}{r|}{\cellcolor[HTML]{AAD18F}21.59} & \multicolumn{1}{r|}{\cellcolor[HTML]{A9D08E}22.19} & \multicolumn{1}{r|}{\cellcolor[HTML]{A9D08E}22.28} & \multicolumn{1}{r|}{\cellcolor[HTML]{ABD190}21.14}      
\\ \hline
34                                        & 78                                       & 2                                        & ZKC                               & \multicolumn{1}{r|}{\cellcolor[HTML]{B7D8A0}12.18} & \multicolumn{1}{r|}{\cellcolor[HTML]{C2DEAF}4.00}     & \multicolumn{1}{r|}{\cellcolor[HTML]{C4DFB1}2.82}  & \multicolumn{1}{r|}{\cellcolor[HTML]{C5E0B3}2.09}  & \multicolumn{1}{r|}{\cellcolor[HTML]{C5E0B3}1.95}  & \multicolumn{1}{r|}{\cellcolor[HTML]{C5E0B3}1.73}  & \multicolumn{1}{r|}{\cellcolor[HTML]{C6E0B4}1.18}  & \multicolumn{1}{r|}{\cellcolor[HTML]{C6E0B4}1.18}  & \multicolumn{1}{r|}{\cellcolor[HTML]{C6E0B4}1.27}  & \multicolumn{1}{r|}{\cellcolor[HTML]{C6E0B4}1.00}     & \multicolumn{1}{r|}{\cellcolor[HTML]{C6E0B4}1}       
\\ \hline
23371                                     & 32832                                    & 350                                      & Twitter                           & \multicolumn{1}{r|}{\cellcolor[HTML]{74A057}37.81} & \multicolumn{1}{r|}{\cellcolor[HTML]{96BF7A}27.83} & \multicolumn{1}{r|}{\cellcolor[HTML]{99C27E}26.80}  & \multicolumn{1}{r|}{\cellcolor[HTML]{99C27E}26.78} & \multicolumn{1}{r|}{\cellcolor[HTML]{ACD292}20.16} & \multicolumn{1}{r|}{\cellcolor[HTML]{99C27E}26.77} & \multicolumn{1}{r|}{\cellcolor[HTML]{A3CB88}23.81} & \multicolumn{1}{r|}{\cellcolor[HTML]{A3CB88}23.80}  & \multicolumn{1}{r|}{\cellcolor[HTML]{A4CB88}23.74} & \multicolumn{1}{r|}{\cellcolor[HTML]{A6CD8B}23.06} & \multicolumn{1}{r|}{\cellcolor[HTML]{ABD190}21.22} 
\\ \hline
1924000                                   & 3298475                                  & 156432                                   & Diggs                             & \multicolumn{1}{r|}{\cellcolor[HTML]{659146}42.49} & \multicolumn{1}{r|}{\cellcolor[HTML]{709C53}39.05} & \multicolumn{1}{r|}{\cellcolor[HTML]{78A35B}36.76} & \multicolumn{1}{r|}{\cellcolor[HTML]{79A45C}36.47} & \multicolumn{1}{r|}{\cellcolor[HTML]{729E55}38.37} & \multicolumn{1}{r|}{\cellcolor[HTML]{709B52}39.21} & \multicolumn{1}{r|}{\cellcolor[HTML]{ACD292}20.11} & \multicolumn{1}{r|}{\cellcolor[HTML]{AED394}18.89} & \multicolumn{1}{r|}{\cellcolor[HTML]{AFD496}17.67} & \multicolumn{1}{r|}{\cellcolor[HTML]{B1D598}16.53} & \multicolumn{1}{r|}{\cellcolor[HTML]{ACD292}19.85} 
\\ \hline
1134891                                   & 2987625                                  & 54983                                    & Youtube                           & \multicolumn{1}{r|}{\cellcolor[HTML]{669348}42.00}    & \multicolumn{1}{r|}{\cellcolor[HTML]{749F56}38.02} & \multicolumn{1}{r|}{\cellcolor[HTML]{7DA860}35.12} & \multicolumn{1}{r|}{\cellcolor[HTML]{85AF69}32.79} & \multicolumn{1}{r|}{\cellcolor[HTML]{86B069}32.59} & \multicolumn{1}{r|}{\cellcolor[HTML]{81AC65}33.92} & \multicolumn{1}{r|}{\cellcolor[HTML]{C3DFB0}3.51}  & \multicolumn{1}{r|}{\cellcolor[HTML]{C4DFB1}2.71}  & \multicolumn{1}{r|}{\cellcolor[HTML]{C5E0B3}1.91}  & \multicolumn{1}{r|}{\cellcolor[HTML]{C6E0B4}1.11}  & \multicolumn{1}{r|}{\cellcolor[HTML]{BFDCAB}6.45} 
\\ \hline
23629                                     & 39195                                    & 75                                       & Ego                               & \multicolumn{1}{r|}{\cellcolor[HTML]{A0C885}24.83} & \multicolumn{1}{r|}{\cellcolor[HTML]{B3D69B}15.34} & \multicolumn{1}{r|}{\cellcolor[HTML]{B4D69C}14.33} & \multicolumn{1}{r|}{\cellcolor[HTML]{B4D69C}14.33} & \multicolumn{1}{r|}{\cellcolor[HTML]{B0D497}17.15} & \multicolumn{1}{r|}{\cellcolor[HTML]{AAD18F}21.81} & \multicolumn{1}{r|}{\cellcolor[HTML]{BBDAA5}9.64}  & \multicolumn{1}{r|}{\cellcolor[HTML]{B9D9A3}10.62} & \multicolumn{1}{r|}{\cellcolor[HTML]{B9D9A2}11.14} & \multicolumn{1}{r|}{\cellcolor[HTML]{BBDAA6}9.05}  & \multicolumn{1}{r|}{\cellcolor[HTML]{BCDAA6}8.89}  
\\ \hline
4658                                      & 33116                                    & 517                                      & LC                                & \multicolumn{1}{r|}{\cellcolor[HTML]{82AC65}33.84} & \multicolumn{1}{r|}{\cellcolor[HTML]{9AC27E}26.62} & \multicolumn{1}{r|}{\cellcolor[HTML]{9DC582}25.61} & \multicolumn{1}{r|}{\cellcolor[HTML]{9DC582}25.61} & \multicolumn{1}{r|}{\cellcolor[HTML]{9EC682}25.52} & \multicolumn{1}{r|}{\cellcolor[HTML]{89B26C}31.81} & \multicolumn{1}{r|}{\cellcolor[HTML]{A8CF8D}22.40}  & \multicolumn{1}{r|}{\cellcolor[HTML]{A5CD8A}23.23} & \multicolumn{1}{r|}{\cellcolor[HTML]{A3CA88}23.98} & \multicolumn{1}{r|}{\cellcolor[HTML]{AAD18F}21.65} & \multicolumn{1}{r|}{\cellcolor[HTML]{A9D08E}22.06} 
\\ \hline
874                                       & 1309                                     & 97                                       & LF                                & \multicolumn{1}{r|}{\cellcolor[HTML]{ADD393}19.29} & \multicolumn{1}{r|}{\cellcolor[HTML]{B9D9A3}10.62} & \multicolumn{1}{r|}{\cellcolor[HTML]{BBDAA5}9.56}  & \multicolumn{1}{r|}{\cellcolor[HTML]{BBDAA5}9.34}  & \multicolumn{1}{r|}{\cellcolor[HTML]{BBDAA6}9.25}  & \multicolumn{1}{r|}{\cellcolor[HTML]{BBDAA5}9.33}  & \multicolumn{1}{r|}{\cellcolor[HTML]{BCDBA7}8.38}  & \multicolumn{1}{r|}{\cellcolor[HTML]{BBDAA5}9.35}  & \multicolumn{1}{r|}{\cellcolor[HTML]{BAD9A4}10.20}  & \multicolumn{1}{r|}{\cellcolor[HTML]{BDDBA8}7.86}  & \multicolumn{1}{r|}{\cellcolor[HTML]{BBDAA6}9.11}  
\\ \hline
1858                                      & 12534                                    & 206                                      & PF                                & \multicolumn{1}{r|}{\cellcolor[HTML]{B9D9A3}10.62} & \multicolumn{1}{r|}{\cellcolor[HTML]{BFDCAA}6.66}  & \multicolumn{1}{r|}{\cellcolor[HTML]{C0DDAC}5.43}  & \multicolumn{1}{r|}{\cellcolor[HTML]{C1DDAD}5.21}  & \multicolumn{1}{r|}{\cellcolor[HTML]{C1DDAD}5.13}  & \multicolumn{1}{r|}{\cellcolor[HTML]{C1DDAD}5.25}  & \multicolumn{1}{r|}{\cellcolor[HTML]{C4DFB1}2.94}  & \multicolumn{1}{r|}{\cellcolor[HTML]{C3DEAF}3.78}  & \multicolumn{1}{r|}{\cellcolor[HTML]{C1DEAE}4.64}  & \multicolumn{1}{r|}{\cellcolor[HTML]{C4DFB2}2.60}   & \multicolumn{1}{r|}{\cellcolor[HTML]{C4DFB1}2.71}    
\\ \hline
22470                                     & 171002                                   & 2643                                     & MFb                               & \multicolumn{1}{r|}{\cellcolor[HTML]{9EC682}25.44} & \multicolumn{1}{r|}{\cellcolor[HTML]{A9D08E}22.16} & \multicolumn{1}{r|}{\cellcolor[HTML]{ABD190}21.11} & \multicolumn{1}{r|}{\cellcolor[HTML]{ABD190}21.11} & \multicolumn{1}{r|}{\cellcolor[HTML]{ABD190}21.10}  & \multicolumn{1}{r|}{\cellcolor[HTML]{ABD190}21.11} & \multicolumn{1}{r|}{\cellcolor[HTML]{B3D69B}15.07} & \multicolumn{1}{r|}{\cellcolor[HTML]{B2D59A}15.80}  & \multicolumn{1}{r|}{\cellcolor[HTML]{A7CE8C}22.70}  & \multicolumn{1}{r|}{\cellcolor[HTML]{ACD291}20.43} & \multicolumn{1}{r|}{\cellcolor[HTML]{B1D498}16.8}   
\\ \hline
54574                                     & 498202                                   & 6420                                     & DHR                               & \multicolumn{1}{r|}{\cellcolor[HTML]{6E9A50}39.77} & \multicolumn{1}{r|}{\cellcolor[HTML]{7CA75F}35.42} & \multicolumn{1}{r|}{\cellcolor[HTML]{84AE67}33.21} & \multicolumn{1}{r|}{\cellcolor[HTML]{88B26B}32.00}    & \multicolumn{1}{r|}{\cellcolor[HTML]{88B26C}31.90}  & \multicolumn{1}{r|}{\cellcolor[HTML]{81AB64}34.2}  & \multicolumn{1}{r|}{\cellcolor[HTML]{BFDCAA}6.78}  & \multicolumn{1}{r|}{\cellcolor[HTML]{BEDCA9}7.26}  & \multicolumn{1}{r|}{\cellcolor[HTML]{BDDBA8}7.73}  & \multicolumn{1}{r|}{\cellcolor[HTML]{C1DDAD}5.21}  & \multicolumn{1}{r|}{\cellcolor[HTML]{C0DDAB}6.01}  
\\ \hline
41774                                     & 125826                                   & 4914                                     & DRO                               & \multicolumn{1}{r|}{\cellcolor[HTML]{659147}42.43} & \multicolumn{1}{r|}{\cellcolor[HTML]{7BA65E}35.74} & \multicolumn{1}{r|}{\cellcolor[HTML]{79A45C}36.42} & \multicolumn{1}{r|}{\cellcolor[HTML]{80AB64}34.22} & \multicolumn{1}{r|}{\cellcolor[HTML]{81AB64}34.12} & \multicolumn{1}{r|}{\cellcolor[HTML]{80AA63}34.45} & \multicolumn{1}{r|}{\cellcolor[HTML]{B5D79E}13.50}  & \multicolumn{1}{r|}{\cellcolor[HTML]{B5D79E}13.40}  & \multicolumn{1}{r|}{\cellcolor[HTML]{B6D79F}13.13} & \multicolumn{1}{r|}{\cellcolor[HTML]{B6D79F}12.94} & \multicolumn{1}{r|}{\cellcolor[HTML]{81AB64}34.18}     
\\ \hline
47539                                     & 222887                                   & 5592                                     & DHU                               & \multicolumn{1}{r|}{\cellcolor[HTML]{5B883C}45.40}  & \multicolumn{1}{r|}{\cellcolor[HTML]{7BA65E}35.77} & \multicolumn{1}{r|}{\cellcolor[HTML]{7FAA62}34.52} & \multicolumn{1}{r|}{\cellcolor[HTML]{80AA63}34.33} & \multicolumn{1}{r|}{\cellcolor[HTML]{81AB64}34.13} & \multicolumn{1}{r|}{\cellcolor[HTML]{719D53}38.85} & \multicolumn{1}{r|}{\cellcolor[HTML]{9BC37F}26.35} & \multicolumn{1}{r|}{\cellcolor[HTML]{99C17D}27.02} & \multicolumn{1}{r|}{\cellcolor[HTML]{9DC581}25.84} & \multicolumn{1}{r|}{\cellcolor[HTML]{9DC582}25.61} & \multicolumn{1}{r|}{\cellcolor[HTML]{9EC683}25.33} 
\\ \hline
37700                                     & 289003                                   & 4435                                     & MG                                & \multicolumn{1}{r|}{\cellcolor[HTML]{8DB670}30.54} & \multicolumn{1}{r|}{\cellcolor[HTML]{97C07B}27.43} & \multicolumn{1}{r|}{\cellcolor[HTML]{9CC480}26.07} & \multicolumn{1}{r|}{\cellcolor[HTML]{9BC380}26.25} & \multicolumn{1}{r|}{\cellcolor[HTML]{9CC480}26.07} & \multicolumn{1}{r|}{\cellcolor[HTML]{9BC37F}26.34} & \multicolumn{1}{r|}{\cellcolor[HTML]{B2D599}16.14} & \multicolumn{1}{r|}{\cellcolor[HTML]{B2D59A}15.49} & \multicolumn{1}{r|}{\cellcolor[HTML]{B2D599}16.05} & \multicolumn{1}{r|}{\cellcolor[HTML]{BAD9A4}10.35} & \multicolumn{1}{r|}{\cellcolor[HTML]{B3D69B}14.86} 
\\ \hline
7624                                      & 27806                                    & 759                                      & L                                 & \multicolumn{1}{r|}{\cellcolor[HTML]{9AC27F}26.55} & \multicolumn{1}{r|}{\cellcolor[HTML]{9FC683}25.25} & \multicolumn{1}{r|}{\cellcolor[HTML]{A3CA87}24.04} & \multicolumn{1}{r|}{\cellcolor[HTML]{A3CB88}23.82} & \multicolumn{1}{r|}{\cellcolor[HTML]{A3CB88}23.79} & \multicolumn{1}{r|}{\cellcolor[HTML]{A3CB88}23.81} & \multicolumn{1}{r|}{\cellcolor[HTML]{AFD395}18.34} & \multicolumn{1}{r|}{\cellcolor[HTML]{AFD396}18.11} & \multicolumn{1}{r|}{\cellcolor[HTML]{AFD496}17.75} & \multicolumn{1}{r|}{\cellcolor[HTML]{AFD496}17.71} & \multicolumn{1}{r|}{\cellcolor[HTML]{AFD496}17.70}  
\\ \hline
50516                                     & 819306                                   & 5943                                     & FbAR                              & \multicolumn{1}{r|}{\cellcolor[HTML]{6D994F}39.97} & \multicolumn{1}{r|}{\cellcolor[HTML]{87B06A}32.40}  & \multicolumn{1}{r|}{\cellcolor[HTML]{8BB46E}31.18} & \multicolumn{1}{r|}{\cellcolor[HTML]{8BB56F}30.95} & \multicolumn{1}{r|}{\cellcolor[HTML]{8BB56F}30.93} & \multicolumn{1}{r|}{\cellcolor[HTML]{8AB46E}31.30}  & \multicolumn{1}{r|}{\cellcolor[HTML]{91BA74}29.43} & \multicolumn{1}{r|}{\cellcolor[HTML]{92BA75}29.14} & \multicolumn{1}{r|}{\cellcolor[HTML]{8CB56F}30.85} & \multicolumn{1}{r|}{\cellcolor[HTML]{93BC77}28.56} & \multicolumn{1}{r|}{\cellcolor[HTML]{91BA75}29.28} 
\\ \hline
13867                                     & 86858                                    & 1383                                     & FbA                               & \multicolumn{1}{r|}{\cellcolor[HTML]{548235}47.29} & \multicolumn{1}{r|}{\cellcolor[HTML]{86B06A}32.45} & \multicolumn{1}{r|}{\cellcolor[HTML]{8BB56F}31.05} & \multicolumn{1}{r|}{\cellcolor[HTML]{8DB670}30.55} & \multicolumn{1}{r|}{\cellcolor[HTML]{6A964C}40.87} & \multicolumn{1}{r|}{\cellcolor[HTML]{59873A}45.89} & \multicolumn{1}{r|}{\cellcolor[HTML]{87B16A}32.28} & \multicolumn{1}{r|}{\cellcolor[HTML]{88B26C}31.83} & \multicolumn{1}{r|}{\cellcolor[HTML]{84AE67}33.28} & \multicolumn{1}{r|}{\cellcolor[HTML]{8DB671}30.46} & \multicolumn{1}{r|}{\cellcolor[HTML]{88B26B}32.01} 
\\ \hline
7058                                      & 89455                                    & 784                                      & FbG                               & \multicolumn{1}{r|}{\cellcolor[HTML]{AAD18F}21.95} & \multicolumn{1}{r|}{\cellcolor[HTML]{ACD292}20.22} & \multicolumn{1}{r|}{\cellcolor[HTML]{AED394}18.93} & \multicolumn{1}{r|}{\cellcolor[HTML]{AED394}18.71} & \multicolumn{1}{r|}{\cellcolor[HTML]{ADD394}19.18} & \multicolumn{1}{r|}{\cellcolor[HTML]{ADD394}19.20}  & \multicolumn{1}{r|}{\cellcolor[HTML]{B5D79D}13.97} & \multicolumn{1}{r|}{\cellcolor[HTML]{B5D79D}13.75} & \multicolumn{1}{r|}{\cellcolor[HTML]{B3D69A}15.39} & \multicolumn{1}{r|}{\cellcolor[HTML]{B6D79F}13.13} & \multicolumn{1}{r|}{\cellcolor[HTML]{B5D79E}13.68}  
\\ \hline
27918                                     & 206259                                   & 3284                                     & FbN                               & \multicolumn{1}{r|}{\cellcolor[HTML]{82AC65}33.82} & \multicolumn{1}{r|}{\cellcolor[HTML]{A6CD8B}23.03} & \multicolumn{1}{r|}{\cellcolor[HTML]{AAD18F}22.00}    & \multicolumn{1}{r|}{\cellcolor[HTML]{AAD18F}21.95} & \multicolumn{1}{r|}{\cellcolor[HTML]{AAD18F}21.96} & \multicolumn{1}{r|}{\cellcolor[HTML]{A9D08E}22.01} & \multicolumn{1}{r|}{\cellcolor[HTML]{B6D79F}12.85} & \multicolumn{1}{r|}{\cellcolor[HTML]{B6D89F}12.64} & \multicolumn{1}{r|}{\cellcolor[HTML]{B7D8A0}12.18} & \multicolumn{1}{r|}{\cellcolor[HTML]{B7D8A0}12.10}  & \multicolumn{1}{r|}{\cellcolor[HTML]{B7D8A0}12.40} 
\\ \hline
5909                                      & 41729                                    & 562                                      & FbP                               & \multicolumn{1}{r|}{\cellcolor[HTML]{89B36C}31.73} & \multicolumn{1}{r|}{\cellcolor[HTML]{A6CE8B}22.90}  & \multicolumn{1}{r|}{\cellcolor[HTML]{AAD18F}21.76} & \multicolumn{1}{r|}{\cellcolor[HTML]{AAD190}21.40}  & \multicolumn{1}{r|}{\cellcolor[HTML]{AAD18F}21.87} & \multicolumn{1}{r|}{\cellcolor[HTML]{AAD18F}21.89} & \multicolumn{1}{r|}{\cellcolor[HTML]{B2D59A}15.89} & \multicolumn{1}{r|}{\cellcolor[HTML]{B3D59A}15.47} & \multicolumn{1}{r|}{\cellcolor[HTML]{B3D69B}15.15} & \multicolumn{1}{r|}{\cellcolor[HTML]{B3D69B}14.90}  & \multicolumn{1}{r|}{\cellcolor[HTML]{B3D69B}15.31} 
\\ \hline
11566                                     & 67114                                    & 1051                                     & FbPF                              & \multicolumn{1}{r|}{\cellcolor[HTML]{6E9A51}39.61} & \multicolumn{1}{r|}{\cellcolor[HTML]{87B16B}32.21} & \multicolumn{1}{r|}{\cellcolor[HTML]{8CB56F}30.85} & \multicolumn{1}{r|}{\cellcolor[HTML]{8DB670}30.57} & \multicolumn{1}{r|}{\cellcolor[HTML]{8DB771}30.39} & \multicolumn{1}{r|}{\cellcolor[HTML]{8DB671}30.48} & \multicolumn{1}{r|}{\cellcolor[HTML]{9BC37F}26.30}  & \multicolumn{1}{r|}{\cellcolor[HTML]{9CC480}26.12} & \multicolumn{1}{r|}{\cellcolor[HTML]{9DC581}25.85} & \multicolumn{1}{r|}{\cellcolor[HTML]{9FC783}25.21} & \multicolumn{1}{r|}{\cellcolor[HTML]{9BC480}26.21} 
\\ \hline
3893                                      & 17262                                    & 387                                      & FbT                               & \multicolumn{1}{r|}{\cellcolor[HTML]{9CC481}25.93} & \multicolumn{1}{r|}{\cellcolor[HTML]{A7CE8C}22.84} & \multicolumn{1}{r|}{\cellcolor[HTML]{A0C885}24.70}  & \multicolumn{1}{r|}{\cellcolor[HTML]{A2C986}24.29} & \multicolumn{1}{r|}{\cellcolor[HTML]{AFD496}17.73} & \multicolumn{1}{r|}{\cellcolor[HTML]{AFD496}17.77} & \multicolumn{1}{r|}{\cellcolor[HTML]{ADD393}19.37} & \multicolumn{1}{r|}{\cellcolor[HTML]{AED395}18.46} & \multicolumn{1}{r|}{\cellcolor[HTML]{B5D79E}13.71} & \multicolumn{1}{r|}{\cellcolor[HTML]{B5D79E}13.36} & \multicolumn{1}{r|}{\cellcolor[HTML]{B0D496}17.63} 
\\ \hline
\end{tabular}
\end{table}

\subsection{Results From Social Networks}
 
The observation that real-world social networks tend to contain dense communities suggests that community based driver node selection would have a significant advantage over global selection. This relationship with density is also apparent in the generated networks. To verify whether this intuition is correct, we conduct similar analysis to this performed on generated networks.  First, we analyse the percentage of nodes influenced by each method over 100 iterations with a seed set size of 20\% of driver nodes. We have run the experiments for the seed set sizes from 1\%, 10\%, 20\%, 30\%, 40\% and 50\%. We show the comparison in case of 20\% seed size, as it is the lowest seed set level to reach maximum influence in at most 100 iterations. We note however that there is also improvements at smaller seed set sizes. 

\begin{figure}[htb]
    \centering
    \includegraphics[scale=0.5]{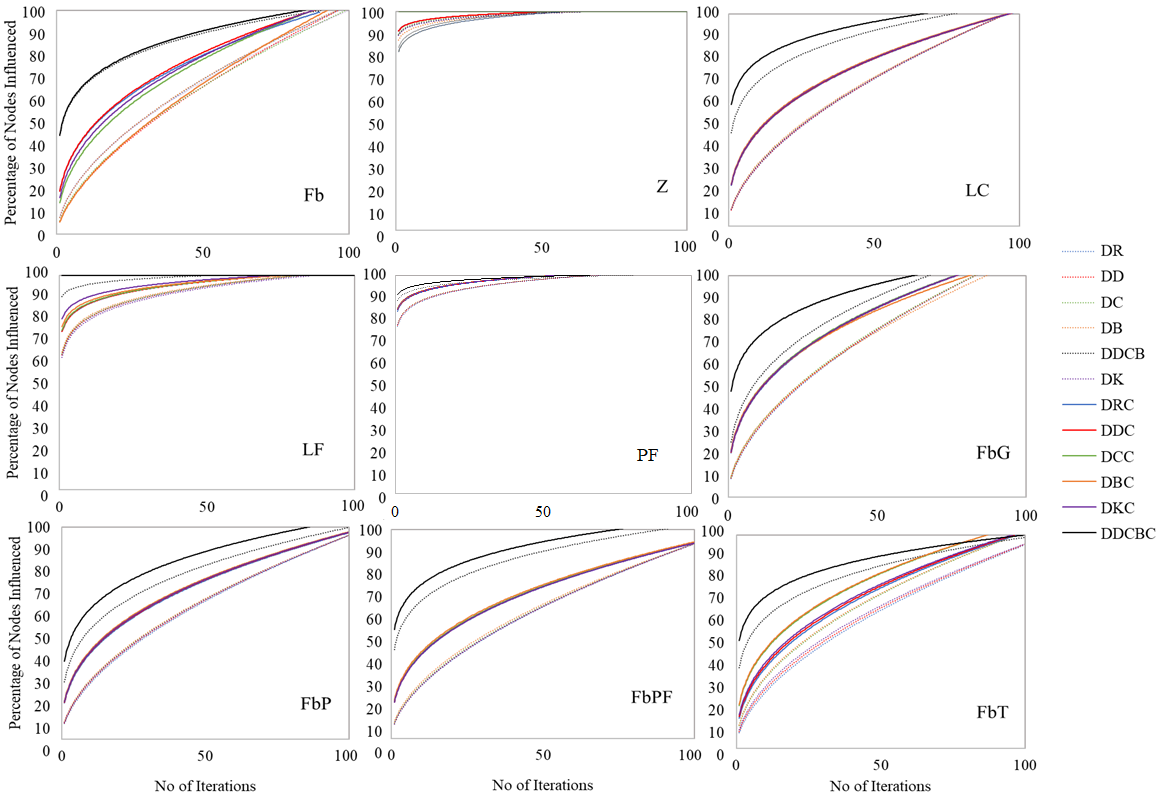}
    \caption{Percentage of Number of Nodes Influenced in FB, Z, LC, LF, PF, FbG, FbP, FbPF and FbT Networks. A Comparison of all methods for 100 iterations.}
    \label{fig:FB-ZKC-LC-LF-PF-FbG-FbP-FbPF-FbT}
\end{figure}

\begin{figure}[htb]
    \centering
    \includegraphics[scale=0.5]{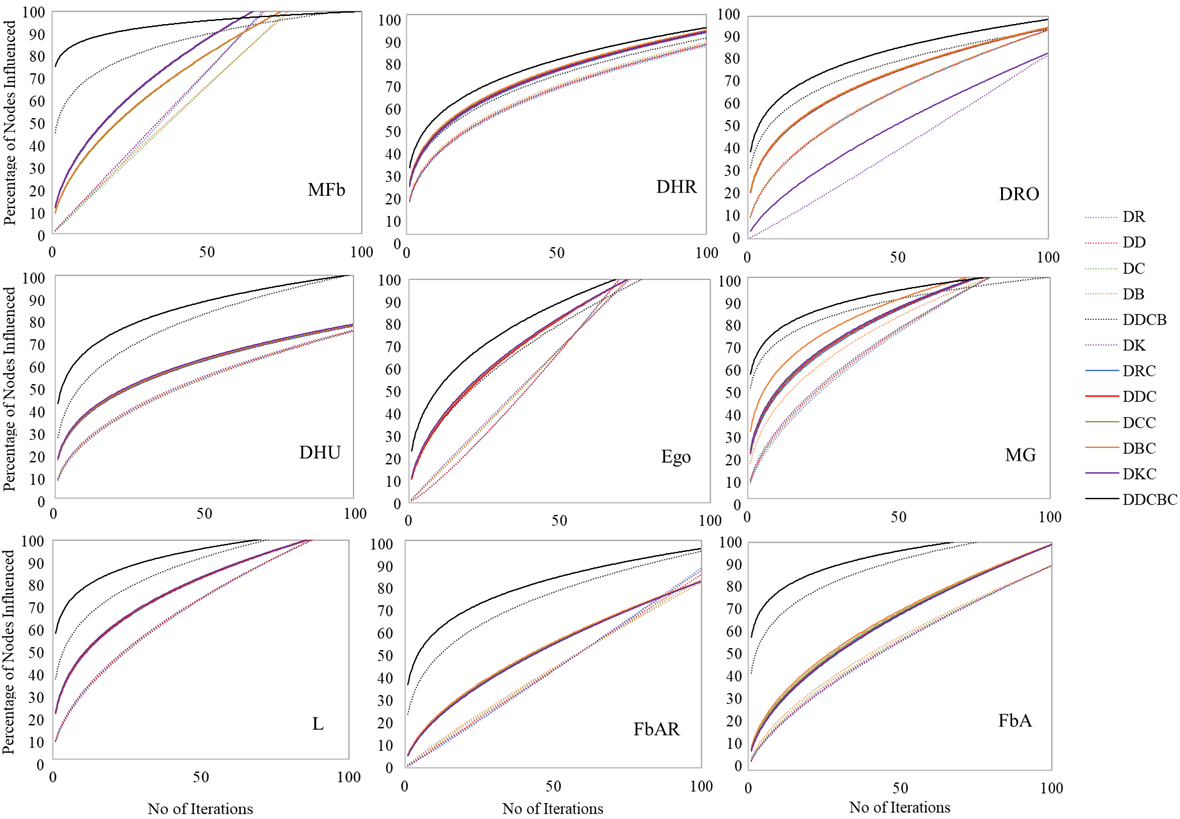}
    \caption{Percentage of Number of Nodes Influenced in MFb, DHR, DRO, DHU, MG, L, FbAR and FbA Networks. A Comparison of all methods for 100 iterations.}
    \label{fig:MFb-DHR-DRO-DHU-MG-L-FbAR-FbA}
\end{figure}

\begin{figure}[htb]
    \centering
    \includegraphics[scale=0.5]{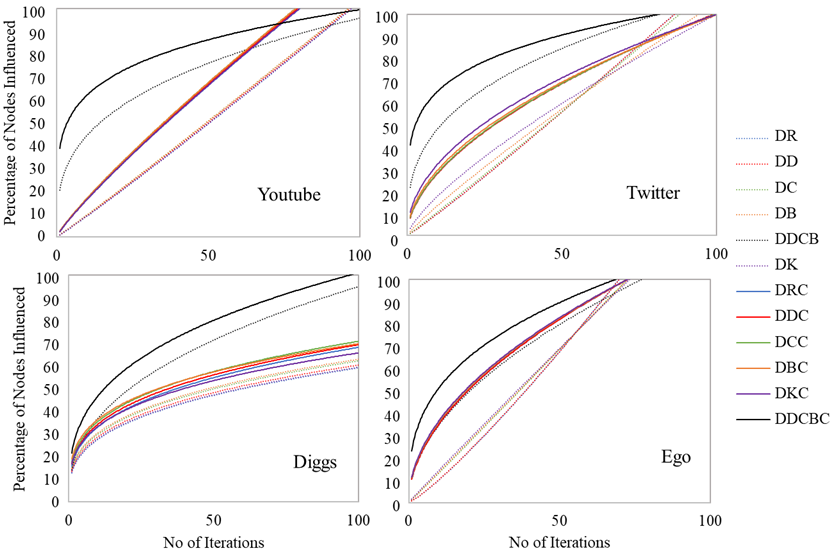}
    \caption{Percentage of Number of Nodes Influenced in Youtube, Twitter, Diggs and Ego Networks. A Comparison of all methods for 100 iterations.}
    \label{fig:Youtube-Twitter-Diggs-Ego}
\end{figure}

\subsubsection{What is the speed and reach of the influence spread?}
Figures~\ref{fig:FB-ZKC-LC-LF-PF-FbG-FbP-FbPF-FbT},~\ref{fig:MFb-DHR-DRO-DHU-MG-L-FbAR-FbA} and~\ref{fig:Youtube-Twitter-Diggs-Ego} show a comparison between global-level driver based seed selection methods and community-level driver based seed selection methods. We grouped the networks on the basis of their sizes and densities to analyse the results effectively.
From Figure~\ref{fig:FB-ZKC-LC-LF-PF-FbG-FbP-FbPF-FbT}, we see a higher density of networks. The densities of these networks are: FB (0.01), Z ( 0.13), LC (0.003), LF (0.003), PF (0.007), FbG (0.003), FbP (0.002), FbPF (0.001) and FbT (0.002).   Overall comparison tells us that, in these networks, there is less difference between the percentage of number of nodes influenced after 100 iterations. Which indicates that when the network's densities are higher, then there is more chance that seed selection methods are able to achieve influence faster. If we look at the Fb network in Figure~\ref{fig:FB-ZKC-LC-LF-PF-FbG-FbP-FbPF-FbT}, its network density is 0.01 which is greater than the rest of the networks except the network Z which has the highest density of 0.14. If we compare the plots, we see that DDCBC method also works exceptionally better in most networks as compared to the rest of the methods. 
From Figure~\ref{fig:MFb-DHR-DRO-DHU-MG-L-FbAR-FbA}, we see the networks with densities ranging from 0.0001 to 0.0009. Densities of these networks are: MFb (0.0006), DHR (0.0003), DRO (0.0001), DHU (0.0001), MG (0.0004), L (0.0009), FbAR (0.0006) and FbA (0.0009). With the lower density networks, we can see that the gain in driver community based methods is more prominent as compared to driver based methods. It means density of the network does play an important role to determine the total number of nodes influenced.
From Figure~\ref{fig:Youtube-Twitter-Diggs-Ego}, we see the networks with the lowest densities ranging from 0.000002 to 0.0001. Densities of these networks are: Youtube (0.000004), Twitter (0.00012), Diggs (0.000002) and Ego (0.00014). In these networks, we see a huge gap between DDCBC method and the rest of the methods. Which means, even in the lowest density networks, when we locally construct communities, the density tend to increase as we can see from Table~\ref{table:Gainoverothermethodssocialnetworks}. Average community density of Youtube was calculated to be 0.000012±0.04, which means if we compare it to the overall network density of 0.000004, it is notably denser. That is why, even in these networks, driver-community based methods specially DDCBC method outperforms the driver based methods.

\subsubsection{How much advantage do community-level driver based seed selection methods give?}
From Table~\ref{table:Gainoverothermethodssocialnetworks}, we see the percentage of gain that DDCBC has over other seed selection methods in terms of number of nodes influenced after 100 iterations when seed size is 20\%. We can see from the table that DDCBC outperforms all methods, but the gain is bigger in terms of global-level driver based methods than the community-level driver based methods. We see this difference in gain mainly because of locally selected and then ranked driver nodes. Also, community creation plays an important role as, the communities are denser than the overall network. From Table~\ref{table:Gainoverothermethodssocialnetworks} we can see that the biggest gain is achieved by DDCBC method over DK method which is 45.89\% in FbA network. And the lowest gain is achieved by DDCBC method over DK method in ZKC network. The reason for lowest or lower gain is that ZKC has the highest network density and smallest size. In denser networks, we tend to see the less gain in DDCBC method. Which precisely can mean that, if we locally identify communities, those have denser structures as compared to the overall network. That is why community-driver based methods combined with ranking of DCB works better than the rest of the methods.
\section{Conclusion and Future Work}\label{conclusion&futurework}

An idea of bringing the methods from control and influence fields together has been proposed in this research. In fact, we played with a research dimension that is at the intersection of both fields and fulfils the objectives of many research questions from both domains.
We proposed, implemented and compared a list of new and novel seed selection methods with the traditional seed selection methods from influence domain and driver seed selection methods from influence meets control field.
In this work, we introduced new seed selection methods, by utilising driver nodes in communities of the networks. The new methods outperformed the old ones. This opens up an avenue in the already existing research of control methods in complex networks. Our community-driver based methods show that, we can achieve maximum influence in fewer number of iterations and with a comparatively less seed set size. Also, if we use ranking mechanisms based upon the centrality measures combining degree, betweenness and closeness, the driver nodes selected as seed nodes perform much better in that case as compared to when we rank them on the basis of individual centrality measures.

Work remains to be done in the context of ranking of driver nodes by using different other algorithms for example, Page Rank, Leader Rank, cluster Rank and K-Shell Decomposition.  
E.g., Page Rank~\cite{brin1998anatomy}, Leader Rank~\cite{lu2011leaders}, Cluster Rank~\cite{chen2013identifying} and K-Shell Decomposition~\cite{liu2015improving}. New methods such as Preferential Matching~\cite{zhang2014structural} can be used to identify driver nodes to improve the efficiency of the seed selection process. 
Another avenue for exploration is the effects of differing influence models, such as the Independent Cascade Model~\cite{duan2009informational}.

\section*{Acknowledgement}
This work was supported in part by the Polish National Science Centre, under Grant no. 2016/21/D/ST6/02408, and in part by the Australian Research Council, Dynamics and Control of Complex Social Networks, under Grant DP190101087.

\bibliographystyle{9-SNMAM}
\bibliography{9-SNMAM.bib}

\begin{thebibliography}{10}
\providecommand{\url}[1]{\texttt{#1}}
\providecommand{\urlprefix}{URL }

\bibitem{borgs2014maximizing}
Borgs, C., Brautbar, M., Chayes, J., Lucier, B.: Maximizing social influence in
  nearly optimal time. In: Proceedings of the twenty-fifth annual ACM-SIAM
  symposium on Discrete algorithms. pp. 946--957. SIAM (2014)

\bibitem{brin1998anatomy}
Brin, S., Page, L.: The anatomy of a large-scale hypertextual web search
  engine. Computer networks and ISDN systems  30(1-7),  107--117 (1998)

\bibitem{chen2013identifying}
Chen, D.B., Gao, H., L{\"u}, L., Zhou, T.: Identifying influential nodes in
  large-scale directed networks: the role of clustering. PloS one  8(10),
  e77455 (2013)

\bibitem{chen2014cim}
Chen, Y.C., Zhu, W.Y., Peng, W.C., Lee, W.C., Lee, S.Y.: Cim: community-based
  influence maximization in social networks. ACM Transactions on Intelligent
  Systems and Technology (TIST)  5(2),  1--31 (2014)

\bibitem{cheng2013staticgreedy}
Cheng, S., Shen, H., Huang, J., Zhang, G., Cheng, X.: Staticgreedy: solving the
  scalability-accuracy dilemma in influence maximization. In: Proceedings of
  the 22nd ACM international conference on Information \& Knowledge Management.
  pp. 509--518 (2013)

\bibitem{cohen2014sketch}
Cohen, E., Delling, D., Pajor, T., Werneck, R.F.: Sketch-based influence
  maximization and computation: Scaling up with guarantees. In: Proceedings of
  the 23rd ACM International Conference on Conference on Information and
  Knowledge Management. pp. 629--638 (2014)

\bibitem{d2016influence}
D'Angelo, G., Severini, L., Velaj, Y.: Influence maximization in the
  independent cascade model. In: ICTCS. pp. 269--274 (2016)

\bibitem{duan2009informational}
Duan, W., Gu, B., Whinston, A.B.: Informational cascades and software adoption
  on the internet: an empirical investigation. MIS quarterly pp. 23--48 (2009)

\bibitem{girvan2002community}
Girvan, M., Newman, M.E.: Community structure in social and biological
  networks. Proceedings of the national academy of sciences  99(12),
  7821--7826 (2002)

\bibitem{goyal2011celf++}
Goyal, A., Lu, W., Lakshmanan, L.V.: Celf++ optimizing the greedy algorithm for
  influence maximization in social networks. In: Proceedings of the 20th
  international conference companion on World wide web. pp. 47--48 (2011)

\bibitem{granovetter1978threshold}
Granovetter, M.: Threshold models of collective behavior. American journal of
  sociology  83(6),  1420--1443 (1978)

\bibitem{kazemzadeh2022influence}
Kazemzadeh, F., Safaei, A.A., Mirzarezaee, M.: Influence maximization in social
  networks using effective community detection. Physica A: Statistical
  Mechanics and its Applications  598,  127314 (2022)

\bibitem{kempe2003maximizing}
Kempe, D., Kleinberg, J., Tardos, {\'E}.: Maximizing the spread of influence
  through a social network. In: Proceedings of the ninth ACM SIGKDD
  international conference on Knowledge discovery and data mining. pp. 137--146
  (2003)

\bibitem{leskovec2005graphs}
Leskovec, J., Kleinberg, J., Faloutsos, C.: Graphs over time: densification
  laws, shrinking diameters and possible explanations. In: Proceedings of the
  eleventh ACM SIGKDD international conference on Knowledge discovery in data
  mining. pp. 177--187 (2005)

\bibitem{liu2015improving}
Liu, Y., Tang, M., Zhou, T., Do, Y.: Improving the accuracy of the k-shell
  method by removing redundant links: From a perspective of spreading dynamics.
  Scientific reports  5(1),  1--11 (2015)

\bibitem{lu2011leaders}
L{\"u}, L., Zhang, Y.C., Yeung, C.H., Zhou, T.: Leaders in social networks, the
  delicious case. PloS one  6(6),  e21202 (2011)

\bibitem{nacher2012dominating}
Nacher, J.C., Akutsu, T.: Dominating scale-free networks with variable scaling
  exponent: heterogeneous networks are not difficult to control. New Journal of
  Physics  14(7),  073005 (2012)

\bibitem{nguyen2016stop}
Nguyen, H.T., Thai, M.T., Dinh, T.N.: Stop-and-stare: Optimal sampling
  algorithms for viral marketing in billion-scale networks. In: Proceedings of
  the 2016 international conference on management of data. pp. 695--710 (2016)

\bibitem{nguyen2017billion}
Nguyen, H.T., Thai, M.T., Dinh, T.N.: A billion-scale approximation algorithm
  for maximizing benefit in viral marketing. IEEE/ACM Transactions On
  Networking  25(4),  2419--2429 (2017)

\bibitem{nguyen2013budgeted}
Nguyen, H., Zheng, R.: On budgeted influence maximization in social networks.
  IEEE Journal on Selected Areas in Communications  31(6),  1084--1094 (2013)

\bibitem{sadaf2021insight}
Sadaf, A., Mathieson, L., Musial, K.: An insight into network structure
  measures and number of driver nodes. In: Proceedings of the 2021 IEEE/ACM
  International Conference on Advances in Social Networks Analysis and Mining.
  pp. 471--478 (2021)

\bibitem{sadaf2022bridge}
Sadaf, Mathieson, B., Musial: A bridge between influence models and control
  methods[manuscript submitted for publication]. IEEE Transactions on Network
  Science and Engineering  (2022)

\bibitem{sanchis2002experimental}
Sanchis, L.A.: Experimental analysis of heuristic algorithms for the dominating
  set problem. Algorithmica  33(1),  3--18 (2002)

\bibitem{sathiyakumari2016community}
Sathiyakumari, K., Vijaya, M.: Community detection based on girvan newman
  algorithm and link analysis of social media. In: Annual Convention of the
  Computer Society of India. pp. 223--234. Springer (2016)

\bibitem{tang2015influence}
Tang, Y., Shi, Y., Xiao, X.: Influence maximization in near-linear time: A
  martingale approach. In: Proceedings of the 2015 ACM SIGMOD international
  conference on management of data. pp. 1539--1554 (2015)

\bibitem{tang2014influence}
Tang, Y., Xiao, X., Shi, Y.: Influence maximization: Near-optimal time
  complexity meets practical efficiency. In: Proceedings of the 2014 ACM SIGMOD
  international conference on Management of data. pp. 75--86 (2014)

\bibitem{zhang2014structural}
Zhang, X., Lv, T., Yang, X., Zhang, B.: Structural controllability of complex
  networks based on preferential matching. PloS one  9(11),  e112039 (2014)

\bibitem{zhu2017emotional}
Zhu, J., Wang, B., Wu, B., Zhang, W.: Emotional community detection in social
  network. IEICE Transactions on Information and Systems  100(10),  2515--2525
  (2017)

\end{thebibliography}

\end{document}